\documentclass[aps,prb,twocolumn,showpacs,preprintnumbers,amsmath,amssymb]{revtex4}

\usepackage[english]{babel}
\usepackage{graphicx}
\usepackage{subfigure}
\usepackage{dcolumn}
\usepackage{bm}
\usepackage{psfrag}
\usepackage{color}
\begin{document}

\preprint{}

\title{Dynamic structure factor for $^3$He in two--dimensions}

\author{M. Nava$^1$, D.E. Galli$^1$, S. Moroni$^2$, and E. Vitali$^1$}
\affiliation{$^1$ Dipartimento di Fisica, Universit\`a degli Studi di Milano, via Celoria 16, 20133 Milano, Italy \\
             $^2$ IOM--CNR DEMOCRITOS National Simulation Center, via Beirut 2-4, 34014 Trieste, Italy}
\date{\today}

\begin{abstract}
Recent neutron scattering experiments on $^3$He films have observed a zero-sound mode,
its dispersion relation and its merging with --and possibly emerging from-- the particle-hole continuum.\cite{nature}
Here we address the study of the excitations in the system via quantum Monte Carlo methods:
we suggest a practical scheme to calculate imaginary time correlation functions 
for moderate-size fermionic systems. Combined with an efficient method for analytic continuation, 
this scheme affords an extremely convincing description of the experimental findings.
\end{abstract}
 

\maketitle
{\em Introduction:} The two isotopes of Helium, $^3$He and $^4$He,
give the opportunity to explore the quantum behavior of many-body systems on a fundamental basis;
at low temperature and pressures, they are the only neutral quantum liquids existing in Nature
and an impressive complexity of physical phenomena is generated by 
merely pair interactions between particles and the effects of quantum statistics.
In the investigation of the fascinating behavior of strongly correlated quantum systems
a key role is naturally played by the low energy dynamics (see, for example, Ref.~\onlinecite{pines}).
In addition, due to the very simple Hamiltonian, $^3$He and $^4$He many--body systems
represent also extremely important reference models and
test cases for general theoretical approaches.\cite{krotscheck2}

Recently inelastic neutron scattering experiments
have been performed on a monolayer of liquid $^3$He adsorbed on suitably preplated graphite:
a collective {\it zero-sound} mode (ZSM) has been detected as a well defined
excitation crossing and possibly reemerging from the particle--hole continuum typical of a Fermi
fluid.\cite{krotscheck1,nature}
From the theoretical side, a quantitative description
of such experimental findings has been achieved by a dynamical 
many body theory, without any adjustable parameters.\cite{nature}
The aim of this work is to undertake an {\it ab--initio} study of the ZSM in a strictly
two--dimensional (2D) $^3$He sample relying on Quantum Monte Carlo (QMC) methods.
It has been shown that this ideal, strictly 2D model offers a realistic representation of the adsorbed liquid layer,
as far as the liquid phase properties are concerned.\cite{whitlock,cepe2d,prbnoi}

The key quantity to be computed to compare with the ZSM observed in neutron scattering experiments 
on $^3$He systems is the coherent dynamic structure factor,\cite{glyde} 
which, apart from kinematical factors, is related to the differential cross section: 
\begin{equation}
\label{sqw}
S(q,\omega)= \frac{1}{2\pi N } \int_{-\infty}^{+\infty}dt\,e^{i\omega t}
\langle e^{i\frac{t}{\hbar} \hat{H}}\,\hat{\rho}_{\vec{q}}\, e^{-i\frac{t}{\hbar} \hat{H}}\,\hat{\rho}_{-\vec{q}} \rangle \quad .
\end{equation}
The brakets indicate a ground state or thermal average, $\hat{H}$ is the Hamiltonian operator, and 
$\hat{\rho}_{\vec{q}} = \sum_{i=1}^{N}\, e^{-i \vec{q} \cdot \vec{\hat{r}}_i}
$ is the local density in Fourier space.
The ZSM of the system manifests itself in the shape of $S(q,\omega)$, appearing
either as sharp peaks if it is long-lived or as broad structures if strong damping is present.\cite{pines}

QMC methods give access to the coherent dynamic structure factor, $S(q,\omega)$, because they allow to
evaluate the intermediate scattering function
$F(q,\tau)= \langle e^{\tau \hat{H}}\,\hat{\rho}_{\vec{q}}\, e^{-\tau \hat{H}}\,\hat{\rho}_{-\vec{q}} \rangle$
by simulating the imaginary time dynamics driven by the Hamiltonian.\cite{ceperley_pp,rqmc}
For a collection of $^3$He atoms, a very accurate microscopic description 
is afforded by the simple Hamiltonian
\begin{equation}
\label{hamiltonian}
\hat{H} = - \frac{\hbar^2}{2m_{3}}\sum_{i=1}^{N} \nabla_{i}^2
+ \sum_{i<j=1}^{N}v\left(\mid\vec{\hat{r}}_i - \vec{\hat{r}}_j\mid\right) \quad .
\end{equation}
where $m_3$ is the mass of $^3$He atoms and $v(r)$ is an effective pair
potential among $^3$He atoms.\cite{Aziz79}

The correlation function $F(q,\tau)$ is the Laplace transform of $S(q,\omega)$. 
Despite the well known difficulties related to the inversion of the Laplace
transform in ill-posed conditions, the evaluation of $S(q,\omega)$ starting
from the QMC estimation of $F(q,\tau)$ \eqref{fqt_qmc} has been proved to 
be fruitful for several bosonic systems.\cite{ceperley,rqmc,gift,gift2,softdisks,gift3}

For a Fermi liquid, the difficulty is further enhanced by the famous {\it sign problem},\cite{kalos}
thereby the computational effort grows exponentially with the imaginary
time and with the number of particles. 
While accurate approximations exist to circumvent this problem in the calculation of static
ground-state properties,\cite{reynolds} we are aware of no applications of approximate 
schemes such as the restricted path~\cite{rpimc} or constrained path~\cite{cpmc}
methods to the calculation of imaginary-time correlation functions.

Focusing on $T=0$ K, QMC calculations of ground state average replaces the unknown exact ground state $\psi_0$
by the imaginary time projection of a trial function $\psi_T$,\cite{rqmc,pigs,patate} $\psi_0 \equiv e^{-\lambda \hat{H}} \psi_T$. The intermediate scattering function then reads:
\begin{equation}
\label{fqt_qmc}
F(q,\tau) = \frac{\langle \psi_T| e^{-\lambda \hat{H}} \, {\rho}_{\vec{q}} \, e^{-\tau \hat{H}}\, 
\hat{\rho}_{-\vec{q}} e^{-\lambda \hat{H}} | \psi_T \rangle}{\langle \psi_T| e^{-(2\lambda+\tau)\hat{H}} |\psi_T \rangle}
\quad .
\end{equation}
Unfortunately, the projection time $\lambda$ required to filter out the exact ground state from the trial function 
is usually larger than the range of $\tau$ needed to extract spectral information,
so that the total imaginary time $2\lambda+\tau$ in Eq. \eqref{fqt_qmc} is too large for practical purposes.
In this paper we propose two related approximations which avoid the extra time $2\lambda$, whereby the calculation
becomes feasible for a few tens $^3$He atoms. The agreement with the measured dynamic structure factor is 
more than satisfactory.

{\em The dynamic fermionic correlation method:} 
We adopt the standard Jastrow-Slater form for the trial function, $\psi_T^F={\mathcal JD}$.
The starting point of the present work is the following approximation:
\begin{equation}
\label{approximation}
\psi_0^F=e^{-\lambda \hat{H}}\psi_T^F \simeq \mathcal{D} e^{-\lambda \hat{H}}{\mathcal J} = \mathcal{D} \psi_0^B
\end{equation}
where $\psi_0^F$ and $\psi_0^B$ are respectively the fermionic and bosonic ground states of the Hamiltonian,
\eqref{hamiltonian}. 
Throughout this paper, a superscript $F(B)$ indicates Fermi(Bose) statistics 
and a subscript $0$ denotes the exact ground state.
The convenience of the approximation \eqref{approximation} is that the extra projection time $\lambda$ 
does not compound the sign problem because it is
applied only to the symmetric factor of $\psi_F^T$.
\begin{figure}[t]
\includegraphics*[scale=0.35]{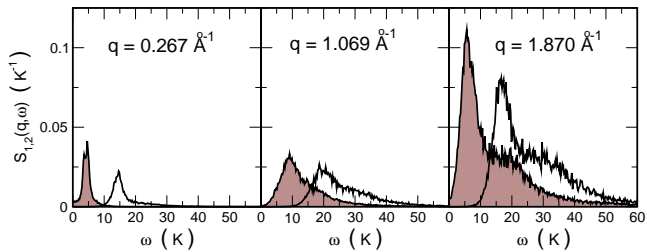}
\caption{(Color online) Comparison between the spectral functions $S_1(q,\omega)$ (filled curve down to the
baseline) obtained with the GIFT algorithm from $F_1(q,\tau)$
and $S_2(q,\omega)$ (unfilled curve) obtained from $F_2(q,\tau)$ with GIFT for
some wave--vectors $q$ at density $\rho=0.047$ \AA$^{-2}$. The two spectral functions have a compatible shape, with a shift in energy
compatible with $E_0^F-E_0^B$.}
\label{fig1}
\end{figure}
In the resulting approximate correlation function 
\begin{equation} \label{sqwvar}
F_{1}(q,\tau) = \frac{\langle \psi_0^{B}|\mathcal{D}^{\star} \, \hat{\rho}_{\vec{q}} \, \, 
e^{-\tau \hat{H}}\, \, \hat{\rho}_{-\vec{q}} \,\mathcal{D}| \psi_0^{B} \rangle}{\langle \psi_0^{B}|\mathcal{D}^{\star} 
e^{-\tau \hat{H}}\,\mathcal{D}| \psi_0^{B}\rangle}
\end{equation}
the projection time between the determinants, which determines the severity of the sign problem,
is limited to $\tau$. $F_1(q,\tau)$ is an approximation of $F(q,\tau)$, and
its inverse Laplace transform, $S_1(q,\omega)$, is an approximation of the coherent dynamic structure factor \eqref{sqw}.
The bias would vanish if $\mathcal{D} \psi_0^B$ were the exact Fermi ground state.
The QMC calculation of $F_1$ requires the ratio of two correlation functions, 
$F_{1}(q,\tau) = F_{2}\left(q,\tau\right)/F_{\mbox{FC}}\left(\tau\right)$, where
\begin{equation}
F_{\mbox{FC}}(\tau) = \frac{\langle \psi_0^B | \mathcal{D}^{\star} e^{-\tau \hat{H}} \mathcal{D} | \psi_0^B \rangle}
{\langle \psi_0^B |e^{-\tau \hat{H}}| \psi_0^B \rangle}
\end{equation}
and
\begin{equation}
\label{ftex}
F_{2}(q,\tau) = \frac{\langle \psi_0^{B}| \, \mathcal{D}^{\star} \,
\hat{\rho}_{\vec{q}} \, \, e^{-\tau \hat{H}}\, \, \hat{\rho}_{-\vec{q}} \,\mathcal{D} | \psi_0^{B} \rangle}
{\langle \psi_0^{B}|e^{-\tau \hat{H}}| \psi_0^{B} \rangle} \quad .
\end{equation}
Both $F_{\mbox{FC}}(\tau)$ and $F_{2}(q,\tau)$ are bosonic correlation functions 
and thus they can be evaluated with great accuracy
by means of exact bosonic QMC methods.
$F_{\mbox{FC}}(\tau)$ is precisely the correlation function that was recently used
in the fermionic correlations (FC) method~\cite{prbnoi}
to study the magnetic properties of $^3$He films.
On the other hand, $F_2$ arises as an extension 
of the FC method to the calculation of the intermediate scattering function,
hence the name ``dynamic fermionic correlation'' (DFC) for the present methodology.
\begin{figure*}[t]
\includegraphics[scale=0.6]{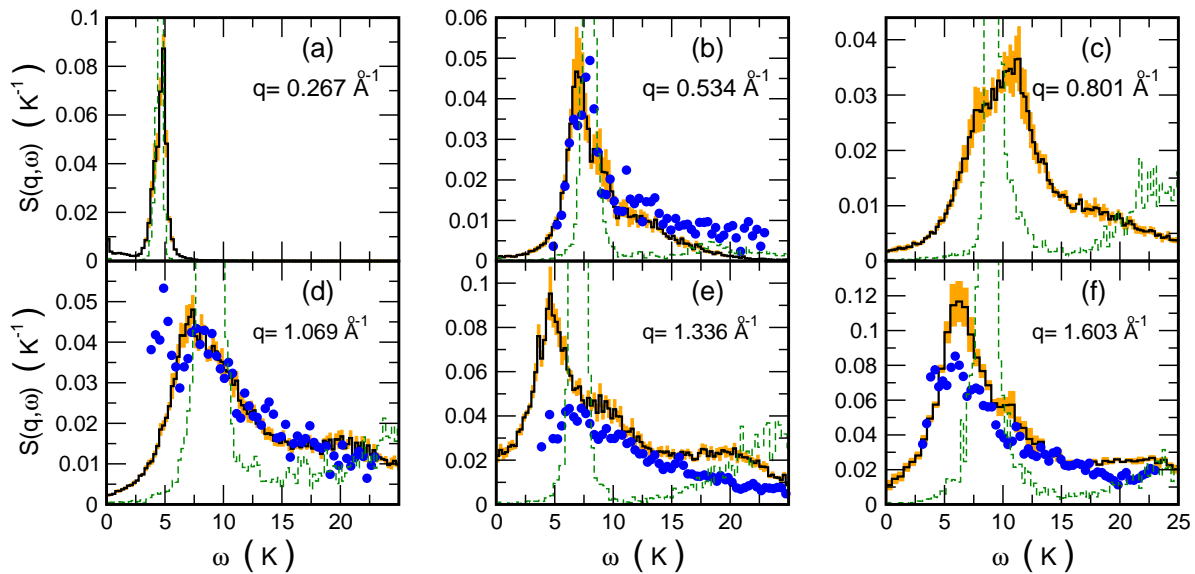}
\caption{(Color online) From left to right the coherent dynamic structure factor,
obtained as an average of several independently extracted $S_1(q,\omega)$,
for increasing wave vectors at $\rho=0.047$ ~\AA$^{-2}$.
The orange shadows represents statistical uncertainties and filled (blue) circles are the available experimental data from
Ref.~\onlinecite{krotscheck1} and Ref.~\onlinecite{nature}.
The wave--vector shown in picture are those accessible from our simulation,
the experimental wave vectors are $q = 0.55$~\AA$^{-1}$(b),
$q = 1.15$~\AA$^{-1}$ (d), $q = 1.25$~\AA$^{-1}$ (e) and $q = 1.65$~\AA$^{-1}$ (f).
We have used different scales in the panels to make more visible the
comparison with experimental data.
The (green) dashed line shows the dynamic structure factor of a fictitious system of bosons of mass $m_3$.
The bosonic peaks in the roton region are $5-9$ times higher than the fermionic ones.
}
\label{fig3}\end{figure*}
Indeed, the function $F_{2}$ possesses very interesting features on its own:
on one hand, it contains the
{\it exact} fermionic spectrum, as can be seen from its spectral resolution:
$F_{2}(q,\tau) = \sum_{n=0}^{+\infty} e^{-\tau \left(E^{F}_n - E_0^B\right)} b_n$,
where $E^{F}_n$ are the fermionic energy eigenvaules, $E_0^B$ is the bosonic ground state energy
and $b_n = |\langle \hat{\rho}_{-\vec{q}} \,\mathcal{D} \,\,\psi_0^{B} | \psi_n^F \rangle|^2 /
\langle \psi_0^{B} | \psi_0^{B} \rangle$.
If, moreover, the approximation \eqref{approximation} is
accurate enough, the coefficients $b_n$ become, apart from an unessential normalization,
the spectral weights of the exact intermediate scattering function, $F(q,\tau)$.
Therefore, computing the inverse Laplace transform of $F_{1}$ and $F_{2}$ we can obtain
two different estimations for the coherent dynamic structure factor, $S_1(q,\omega)$
and $S_2(q,\omega+E_0^F-E_0^B)$ respectively, where the shift in $\omega$ comes from
the definition of $F_{2}$ in terms of the Bose ground state.
\begin{figure*}[t]
\includegraphics[scale=0.62]{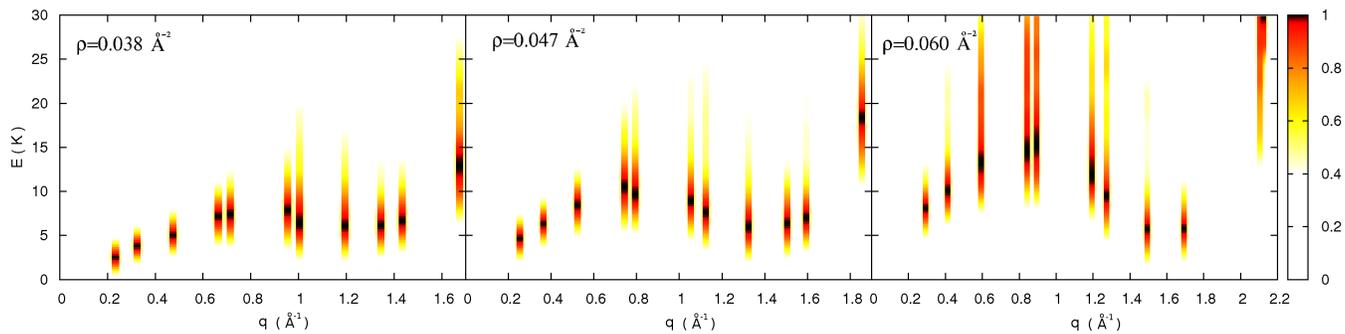}
\caption{(Color online) Color map of normalized $S_2(q,\omega)$ for many wave vector $q$.
For better visibility, each $S_2(q,\omega)$ for different $q$ has been normalized
in order to have their maximum value equal to 1.
The vertical scale have been shifted by a quantity $E_0^B - E_0^F$, so that
the excitation energies are measured with respect to the fermionic ground state.
}
\label{fig4}\end{figure*}
A robust test for the validity of the approximation \eqref{approximation}
is at hand if it turns out that $S_1(q,\omega) \simeq S_2(q,\omega+E_0^F-E_0^B)$:
as already noticed, $F_{2}$ decays with the exact fermionic excitation energies (once shifted);
moreover, if $\mathcal{D} \psi_0^B$ has a small overlap on the fermionic excited states, it follows that
$e^{-\tau \hat{H}} \mathcal{D} \psi_0^B$ will quickly converge in $\tau$ to $e^{-\tau E_0^F} \psi_0^F$.
Therefore $F_{1}(q,\tau) \simeq e^{\tau E_0^F} F_{2}(q,\tau)$, apart from an unessential normalization.
We have indeed verified that, in the present case, $S_1(q,\omega)$ and $S_2(q,\omega+E_0^F-E_0^B)$
possess a very similar shape (see Fig.\ref{fig1}).

{\em Results:} We studied a system of $N=26$ structureless $1/2$-spin fermions
of mass $m_3$, interacting via the Aziz potential,\cite{Aziz79} enclosed in a square box
with periodic boundary conditions. We found in Ref.~\onlinecite{prbnoi} that
this system size offers a good compromise between finite-size effects and computational cost.
Indeed the inverse Laplace transform becomes increasingly difficult as the range of fermionic energy
eigenvalues relevant for the spectral reconstruction departs from the reference energy of the underlying
simulation, which is the bosonic ground state $E_0^B$: this is precisely what happens as
the system size increases because the gap $E_0^F-E_0^B$ is an extensive quantity.
The trial function $\psi_T^F={\mathcal JD}$ is the same as in Ref.~\onlinecite{prbnoi}, namely
a two-body Jastrow factor ${\mathcal J}$  and a Slater determinant ${\mathcal D}$ of plane waves with
simple backflow correlations.
We have focused on a density around $0.047$ \AA$^{-2}$, close to the
experimental conditions.\cite{krotscheck1}
Moreover, we have explored the behavior of the sample at the densities $0.038$ and
$0.060$ \AA$^{-2}$ in order to investigate the density-dependence of the excitations
of the system. In particular, the highest density was chosen very close to the freezing point.\cite{prbnoi}
The QMC evaluation of $F_2$ requires a simple generalization of the methodology we have
followed in Ref.~\onlinecite{prbnoi} to compute $F_{\mbox{FC}}$: a fictitious system of {\it{bosons}} of mass $m_3$ is 
simulated by an {\it{exact}} projector Monte Carlo technique,
the Shadow Path Integral Ground State.\cite{spigs}
The imaginary--time propagation was 1.3125K$^{-1}$ and the density matrix approximation was a Pair Product\cite{ceperley_pp}
with imaginary--time--step of $1/160$ K$^{-1}$.
It is well known that, in order to extract information from imaginary time correlation function,
an inversion of the Laplace transform in ill-posed contitions is necessary.
This can be carried out by means of the Genetic Inversion via Falsification of Theories (GIFT)
,\cite{gift} which has been used to retrieve nontrivial spectral features in the study of low energy excitations
of Bose superfluids\cite{gift,gift2,gift3} and supersolids.\cite{softdisks}

In Fig.~\ref{fig3} we show the comparison between our estimation of the dynamic structure
factor of the $^3$He film and the experimental data.\cite{krotscheck1,nature} 
The dynamic structure factor has been obtained as an average
over several GIFT reconstructions of $S_1(q,\omega)$ from independent estimates of $F_1(q,\tau)$;
this has made possible an estimation of the statistical uncertainties which we have shown in Fig.~\ref{fig3}
with a yellow shadow.
We note that the available experimental wave--vectors do not exactly match the reciprocal space grid
defined by the simulation box.
For $q=0.534$ and $q=1.603$~\AA$^{-1}$, where the mismatch is minimal, a direct comparison is possible and
the agreement is impressive. Inspection of the wave--vector dependence of the spectra shows that the 
discrepancies seen at $q=1.069$ and $q=1.336$ are mostly due to the differences of
$q$ values between theory and experiment.
A major feature of the measured $S(q,\omega)$,
captured also by the dynamical many
body theory in Ref.~\onlinecite{nature},
 is the appearence of a low-energy peak for both small and large 
wavevector, interpreted in Refs.\onlinecite{krotscheck1,nature} as a well defined collective mode,
broadened in the intermediate $q$ range because of mixing with the particle-hole continuum.
In further agreement with the measurements, we find a similar behavior. Indeed the simulation can
provide information even at small wave vectors, not accessible to the experimental probe: at 
$q=0.267$ the collective excitation (ZSM) is most pronounced, and the spectral weight of the
particle-hole is negligible. It is remarkable that both the position and the shape of the 
calculated spectra have a physical meaning and are not artifacts of the reconstruction procedure.
Further support to this conclusion is offered from a comparison with the dynamic structure
factor of the fictitious $^3$He-mass bosonic system. The bosonic spectrum has a completely different
behavior, featuring an extremely sharp peak with the usual phonon-maxon-roton dispersion,
showing that the broadening of the fermionic spectrum is actually related to Fermi statistics.

In Fig.\ref{fig4} we report, in a color plot, the estimated $S_{2}(q,\omega)$. In agreement
with the behavior of $S_1$ shown in Fig.\ref{fig3}, at low $q$ we
find well defined excitation energies; as the wave vector increases, again
we observe broadening of the ZSM.
The ZSM dispersion $E(q)$ as a function of the wave--vector $q$ can be inferred from the distance between the maximum
in $S_2(q,\omega)$ and the value of the energy gap between the Fermionic and the Bosonic ground state.
The dispersion of the ZSM recalls the classical phonon--maxon--roton mode in bulk superfluid $^4$He;
the roton energies decreases with the density and the minimum moves to higher wave--vectors similarly to what
happen for the bosonic liquid. Maxon energies increases with the density and also the zero sound velocity has the same trend.

\begin{figure}[t!]
\includegraphics*[scale=0.35]{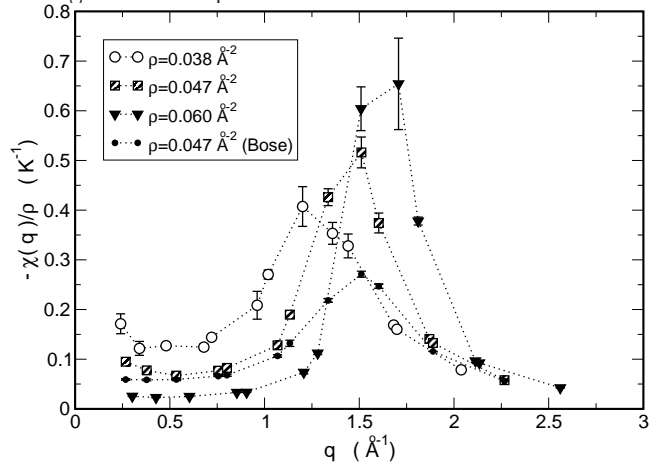}
\caption{The static response function of $^3$He obtained from 
$\chi_{q} = -2\rho\int d\omega \:\frac{S\left(q,\omega\right)}{\omega}$.
(Circles) $\rho=0.038$ ~\AA$^{-2}$;
(Squares) $\rho=0.047$ ~\AA$^{-2}$; (Triangles) $\rho=0.060$ ~\AA$^{-2}$.
The filled circles show the results for a fictitious system of bosons of mass $m_3$ at $\rho=0.047$ ~\AA$^{-2}$.
}
\label{fig6}\end{figure}

As a byproduct of the calculation of $S\left(q,\omega\right)$ we also obtain
the static density response function $\chi(q)$
from the moment -1, shown in Fig.~\ref{fig6}.
We are unaware of previous QMC results for $\chi(q)$.
We have also calculated the static response of the fictitious bosonic
system, which is significantly less structured than the fermionic couterpart, as expected (see Fig.~\ref{fig6}). 

{\em Conclusions:} In this work we have presented an {\it ab--initio}
estimation of the coherent dynamic structure factor of 2D liquid $^3$He, 
a strongly interacting Fermi liquid, combining unbiased QMC sampling techniques
with a statistical method\cite{gift}
for the analytical continuation from imaginary time to real frequencies
under the only approximation \eqref{approximation} 
($\psi_0^F \simeq \mathcal{D} \psi_0^B$).

We find a well defined collective mode (the zero-sound mode) at small wavevector;
its dispersion relation follows a phonon-maxon-roton pattern, with a significant
broadening in the intermediate wavevector range due the mixing with the particle-hole
continuum. These features, including the shape and width of the spectra, are in close agreement 
with a recent neutron scattering experiment,\cite{krotscheck1,nature}
and a sophisticated dynamical many body theory.\cite{nature} 

We expect that the approximation \eqref{approximation}, as suggested by the accuracy of our results 
for 2D liquid $^3$He (See Fig.~\ref{fig1}), is a good one whenever the effects of quantum
statistics are dominated by strong interactions.
The formalism can be readily generalized by replacing the density fluctuations operator $\rho_{\vec q}$ 
to access other spectral properties such as spin fluctuations or particle-hole excitations.

{\em Acknowledgements:} This work has been supported by CASPUR, and by Regione Lombardia and CINECA Consortium through a LISA
Initiative (Laboratory for Interdisciplinary Advanced Simulation) 2012 grant [http://lisa.cilea.it],
and by a grant ``Dote ricerca'': FSE, Regione Lombardia.

\end{document}